# A Gradient Descent Technique Coupled with a Dynamic Simulation to Determine the Near Optimum Orientation of Floor Plan Designs

Eugénio Rodrigues[#1], Adélio Rodrigues Gaspar[#2], Álvaro Gomes[*3]

[#]*ADAI–LAETA, Department of Mechanical Engineering, University of Coimbra*
*Rua Luís Reis Santos, Pólo II, 3030-788 Coimbra, Portugal*

[1]eugenio.rodrigues@gmail.com
[2]adelio.gaspar@dem.uc.pt

[*] *INESCC, Department of Electrical and Computer Engineering, University of Coimbra*
*Rua Luís Reis Santos, Pólo II, 3030-788 Coimbra, Portugal*

[3]alvaro.gomes@uc.pt

**Abstract**

*A prototype tool to assist architects during the early design stage of floor plans has been developed, consisting of an Evolutionary Program for the Space Allocation Problem (EPSAP), which generates sets of floor plan alternatives according to the architect's preferences; and a Floor Plan Performance Optimization Program (FPOP), which optimizes the selected solutions according to thermal performance criteria. The design variables subject to optimization are window position and size, overhangs, fins, wall positioning, and building orientation.*

*A procedure using a transformation operator with gradient descent, such as behavior, coupled with a dynamic simulation engine was developed for the thermal evaluation and optimization process. However, the need to evaluate all possible alternatives regarding designing variables being used during the optimization process leads to an intensive use of thermal simulation, which dramatically increases the simulation time, rendering it unpractical. An alternative approach is a smart optimization approach, which utilizes an oriented and adaptive search technique to efficiently find the near optimum solution.*

*This paper presents the search methodology for the building orientation of floor plan designs, and the corresponding efficiency and effectiveness indicators. The calculations are based on 100 floor plan designs generated by EPSAP. All floor plans have the same design program, location, and weather data, changing only their geometry.*

*Dynamic simulation of buildings was effectively used together with the optimization procedure in this approach to significantly improve the designs. The use of the orientation variable has been included in the algorithm.*

## 1. Introduction

An optimization procedure coupled with a building dynamic simulation has traditionally encountered huge obstacles due to the computational time that is required [1]. However, certain researchers have developed approaches to overcome this difficulty, for example, Wang, Zmeureanu & Rivard [2] presented a multi-objective genetic algorithm for building design optimization coupled with dynamic simulation. To overcome the problem of the necessary time to compute all the dynamic simulations, the methodology was limited to one building design and only a few design variables, such as building orientation, aspect ratio, window type, window-to-wall ratio, wall type, and roof type. The methodology ignored possible arrangements for the interior spaces and acknowledged the whole building as a single thermal zone. In another approach, implemented by Peippo, Lund & Vartiainen [3], a numerical multivariate optimization was employed. The procedure included building orientation, building geometry (aspect ratio), thermal insulation, collector slope, thermal mass, window type and area, collector type, PV capability, energy storage volume, lightning type, lighting control, and exhaust air heat recovery as design variables. The objective was to find the economic optimum according to an energy consumption target. Haase & Amato [4] determined the extent to which building location, climate and orientation contributed to the thermal comfort of a building design. This approach helped to determine the potential improvement of the design.

It may be evident that in order to overcome the problem of computational resources and time needed to carry out a dynamic simulation, the number of designs under optimization (population) should be limited or a very abstract simulation model (fewer design variables) should be used for a faster optimization procedure.

The building orientation plays a significant role in the overall thermal performance of a building. Figure 1 illustrates three different building performance curves for a floor plan design according to their angle of orientation. The bottom axis represents the orientation angle, the top axis, the cardinal direction, and the vertical axis, the amplitude of the curve. Every curve is the result of exhaustively determining the building's thermal discomfort performance for every orientation angle. Each point is the result of the sum of the degree-hours of discomfort for all the spaces. As may be observed, the curves differ greatly depending on several factors such as window position and size, shape of the floor plan, self-projected shading, space volume, thermal properties of the envelop elements, among others.

This procedure is very time consuming due to the necessity of carrying out a dynamic simulation. Another drawback is that determining the minimum value in the curve does not necessarily guarantee that this will be the best orientation when the thermal optimization for the other design variables has ended. The re-positioning of the windows and their sizing will change the curve profile and consequently may shift the minimum point to another position (angle). As each design variable - location, weather data, constructive system, to name but a few, may modify the curve profile, rendering it unique, it may be considered as the fingerprint of the building.

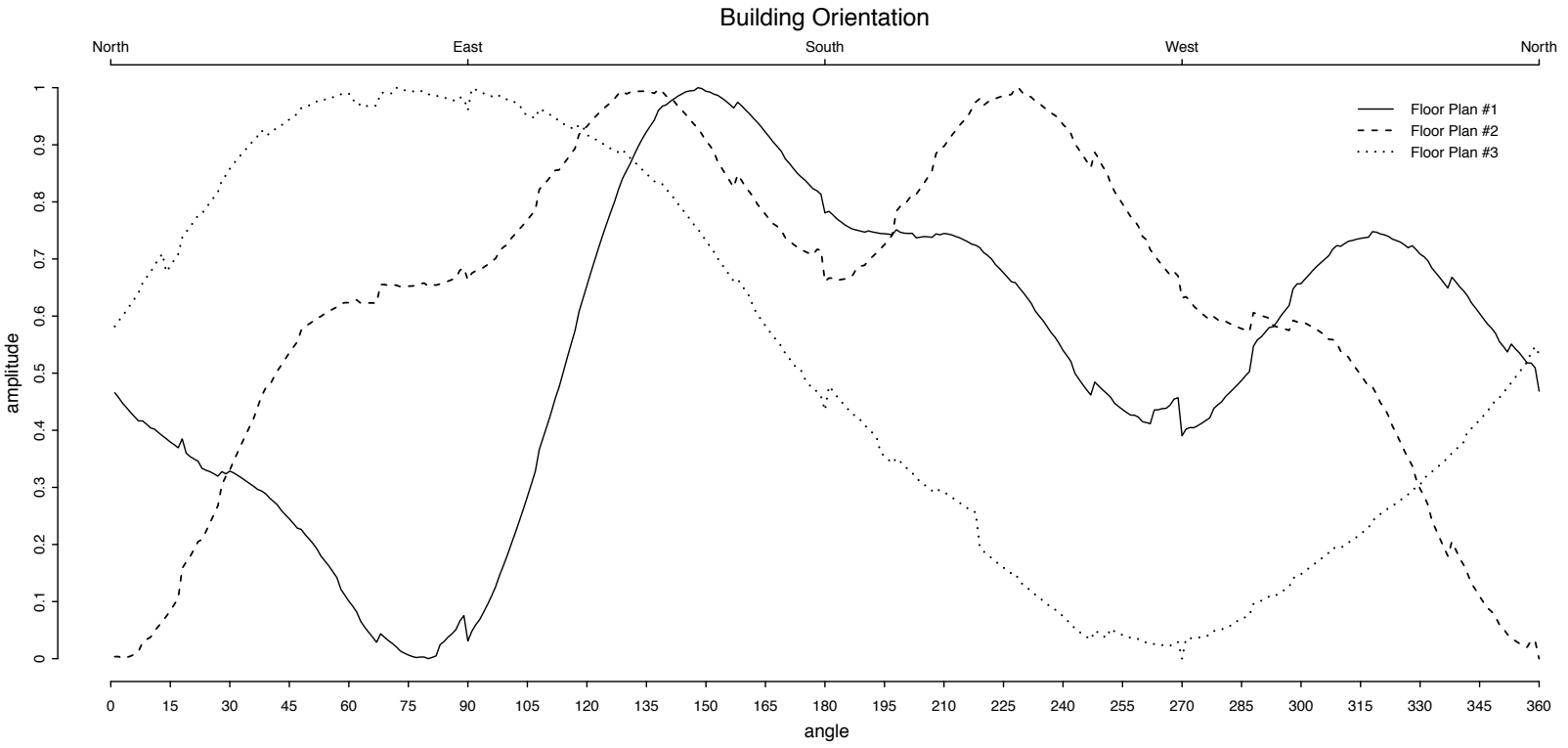

Figure 1: Three different floor plan designs (Floor Plan #1: One level three bedroom house, Floor Plan #2: Two level three bedroom house, and Floor Plan #3: One level with two flats).

The researching team has developed a prototype tool to assist architects in the early design stage of floor plans. It is made up of two modules, the first module is an Evolutionary Program for the Space Allocation Problem module (EPSAP) [5-7] which generates sets of different floor plan design alternatives according to the architect's preferences and constraints. The second module is a Floor Plan Performance Optimization Program module (FPOP) [5], which improves each chosen solution according to thermal performance criteria. The design variables subject to optimization in the latter module are window position, window size, overhangs, fins, interior and exterior wall positioning, and finally, the building orientation.

To overcome the problem associated to the time needed to carry out an exhaustive enumeration using dynamic simulation, a type of gradient descent technique was used as a transformation operator within the optimization procedure to find a better building orientation within a margin of 2% of the curve amplitude in relation to the global optimum. This signifies that on average all floor plan design errors must be below this threshold.

Therefore, considering a problem consisting of 100 detailed floor plan designs, in which one of the design variables to be optimized is the building orientation, this paper presents the methodology that was used in the building orientation transformation operator, the results of the efficiency and effectiveness test, and the main conclusions.

## 2. Methodology

There are many design variables, which may influence the identification of the most suitable building orientation. Practitioner's preferences, constraints (minimum window size, window orientation, window-space area ratio, etc.), the size of the walls and windows, the building system that was used, are a few examples of these variables. The energy assessment of a building requires the calculation of all possible combinations of all design variables. This may become an impractical task due to the required computational effort to carry out the dynamic simulation for all possible combinations.

There are some rule of thumb principles and empirical knowledge in relation to building orientation, which may be applied to potentially identify the most adequate orientation (angle). However, the "best" orientation is dependent on other design variables, which means that a detailed study should be carried out for each set of (other) design variables. Accordingly, a set of different design alternatives, with the same design program and user preferences, may be subjected to an optimization procedure to predict their performance in relation to the thermal performance criteria.

To avoid the previously mentioned problem - computational effort/time - and the exhaustive enumeration of every building orientation angle, a type of gradient descent technique which would change the variable until it reached its near optimum value was employed as a transformation operator. This transformation operator is invoked twice at different moments during the optimization procedure for each floor plan design.

The FPOP optimization procedure consists of several sequential transformation adaptive operators oriented to specific design variables where the building orientation is one of those variables. The sequence of transformation operators that were used included a window translation operator, a building orientation operator, a window dimensioning operator, a wall translation operator, a building orientation operator (second time), a window's overhang operator, and finally a window's fins operator. The building orientation operator is invoked at specific times during the optimization procedure. After the re-positioning of the windows in relation to the user's predetermined orientation, and after the windows have been dimensioned and the walls have been repositioned. The last operators place the overhangs and fins according to the requirements and final orientation of each space. This sequence of operators has proven to be the most effective during the algorithm development.

The assessment of each design is completed using a cost function, which penalizes the thermal discomfort measured in degree-hours for each space (see Equation 1). The limits of thermal comfort conform to the European Standard EN 15251 (2007) operative temperatures for buildings without mechanical cooling systems [8]. When the outdoor mean temperature is below 10ºC or above 30ºC, the algorithm will use those values to calculate the adaptive operative temperature limits. The location and weather data for all designs was for Coimbra, Portugal.

$$f(I) = \sum_{i=1}^{N_s} \sum_{t=1}^{N_t} f_{df}\Big(T_i(t), T_1, T_2\Big) \tag{1}$$

The hourly indoor air temperature results of each space, obtained by the dynamic simulation, were then compared to the thermal comfort interval temperatures. If the indoor temperature at a certain hour is outside the interval, the cost function will penalize the design according to the deviation from the comfort interval ($T_1$ for the lower limit and $T_2$ for the upper limit), multiplied by a corresponding weighting value $w_1$ for the lower limit and $w_2$ for the upper limit (see Equation 2). In Equation 1, the $N_s$ is the number of spaces on the floor plan, the $N_t$ is the number of hours in the year, $T_i(t)$ is the air temperature at the hour $t$ for the space $i$, and $f_{df}$ is the function that calculates the difference between the hourly air temperature $T_i(t)$ and the thermal comfort limits according to Equation 2. The weight difference will allow the user to specify their preferences for future cooling or heating systems, if necessary.

$$f_{df}(T, T_1, T_2) = \begin{cases} w_1(T_1 - T) & \text{if } T_1 > T \\ w_2(T - T_2) & \text{if } T > T_2 \\ 0 & \text{otherwise} \end{cases} \tag{2}$$

In order to determine a suitable building orientation, the transformation operator is a sort of gradient descent technique, which descends through the building orientation curve (see Figure 1) from the initial orientation until it reaches the lowest point possible (which corresponds to a better building orientation for the set of design variables that have been considered). Each step of the approach starts with the value of +30 degrees. This signifies that, for instance, starting from the orientation angle 90 (East) it will determine if the angle 120 has lower penalties. If this is correct, it continues with the same direction. When it fails, the step becomes negative and decreases in size, changing the direction of the search. Subsequently it continues until it no further improvements can be found. When the termination condition of this operator is reached the next operator is activated.

Due to the irregularities of the building orientation curve (see Figure 1) and the benefit of the improvements, it was considered that the admissible penalties deviations would be 2% of the penalties amplitude curve (difference between maximum penalties and the minimum penalties points) at this early design stage. This indicator was chosen instead of the angle deviation indicator since the former could not be representative of the algorithm performance. For example, the orientation curve of some buildings could have two or more local minimums with similar values however with greater angle differences. This occurs with floor plans with one or more axis of geometric symmetry.

## 3. Results and Discussion

A problem test was conducted to determine the effectiveness and efficiency of the building orientation operator. A set of 100 floor plan designs generated by the EPSAP module [5-7] were used. Every floor plan had the same design program, one hall, corridor, kitchen, living room, dinning room, two bathrooms, and three bedrooms. The topological requirements were the same for all designs, and the initial window size was 10% of the corresponding space area in each design solution (see Figure 2 for a floor plan example).

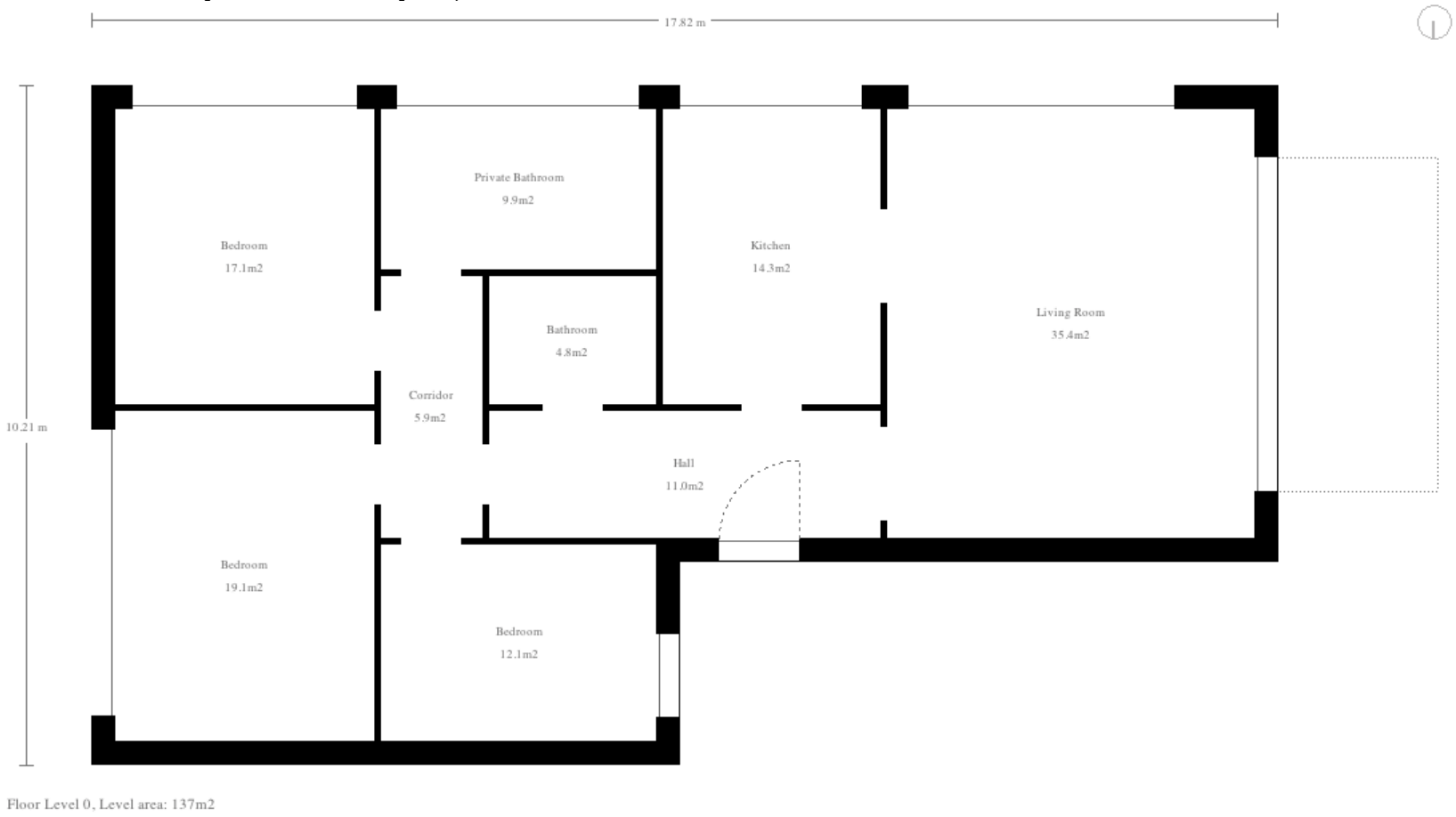

Figure 2: The fittest individual of the FPOP module optimization.

The test was carried out with a dual-core computer at 3.2GHz with 1010MiB of RAM. The algorithm was built in JAVA programming language and was implemented to take full advantage of parallel computing by using two threads. The optimization procedure took 2 days, 13 hours and 33 minutes to conclude. Each simulation took approximately 28 seconds and from a total of 7,833 simulations, the building orientation operator carried out 993 simulations. This is nearly 2.7% of the simulations in comparison to an exhaustive enumeration. This signifies that on average only 4 to 5 steps of the operator are necessary to find the suitable orientation each time it is invoked.

As one may observe on the right side of Figure 3, the building orientation operator significantly impacts the improvement of the building's thermal performance. The relative contribution of this operator, compared to the overall improvement of the floor plan design, is up to 14%, as illustrated in left side of Figure 3.

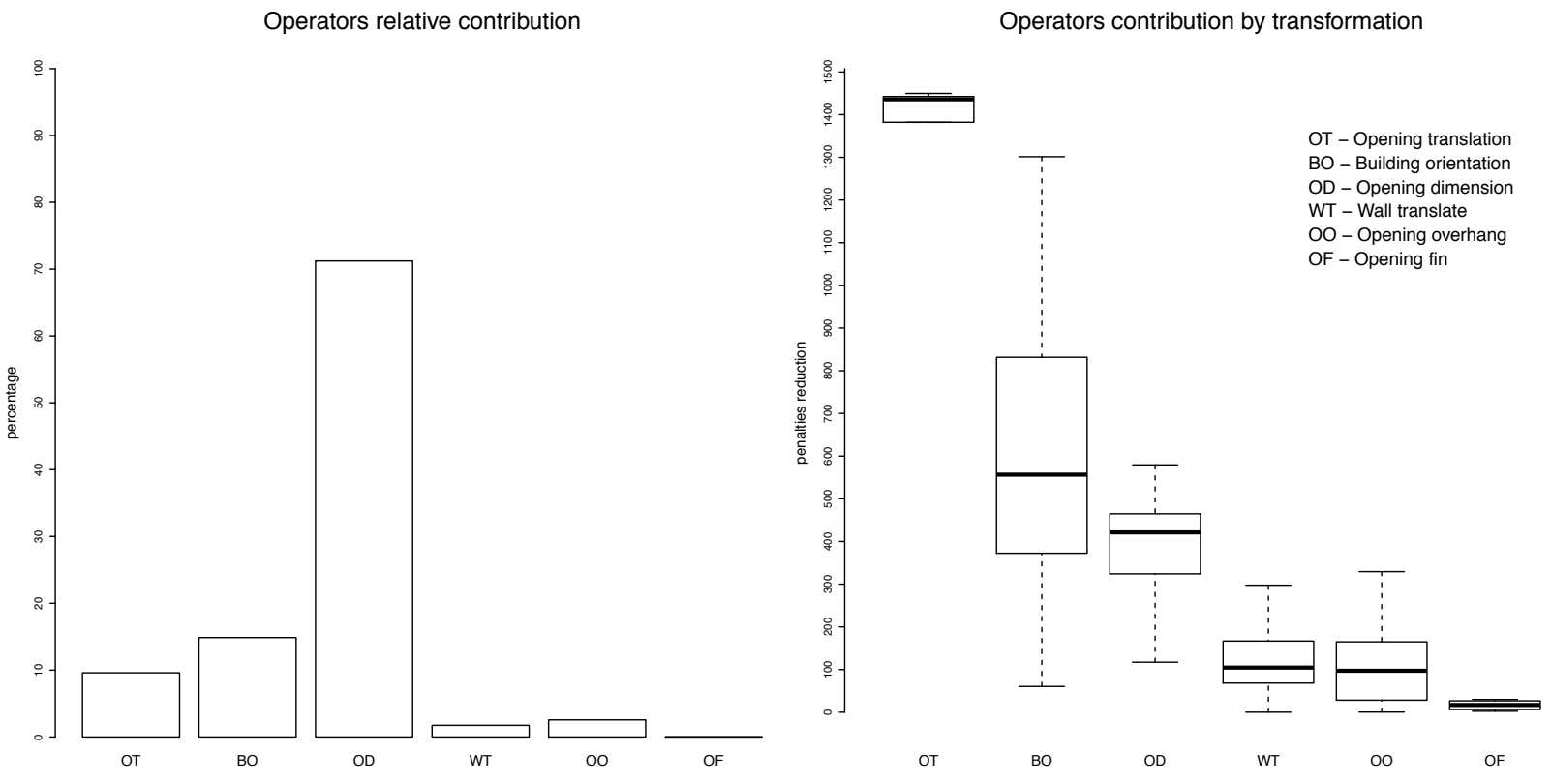

Figure 3: Operators' relative contribution to the total improvement (left side) and operators' contribution by transformation (right side).

The building orientation curves of the optimized individuals differ in their profile. These curves are depicted in Figure 4 and the global optimum (black dot) and determined optimization orientation values (red cross) are marked. If the curves are adjusted to have their maximum value aligned to the left side of the graph (see Figure 5) it is possible to observe the relative position of the global minimums and the best optimization results. The minimum value is usually located in the opposite area of the maximum value within an interval of -90 to +90 degrees. This indicates that, after the optimization procedure is concluded, the curve amplitude increases due to the floor plan design being improved specifically for that location and weather data. For instance, the re-positioning of the windows facing South, West, and East will augment the penalties when the building is oriented with the same façade to the North.

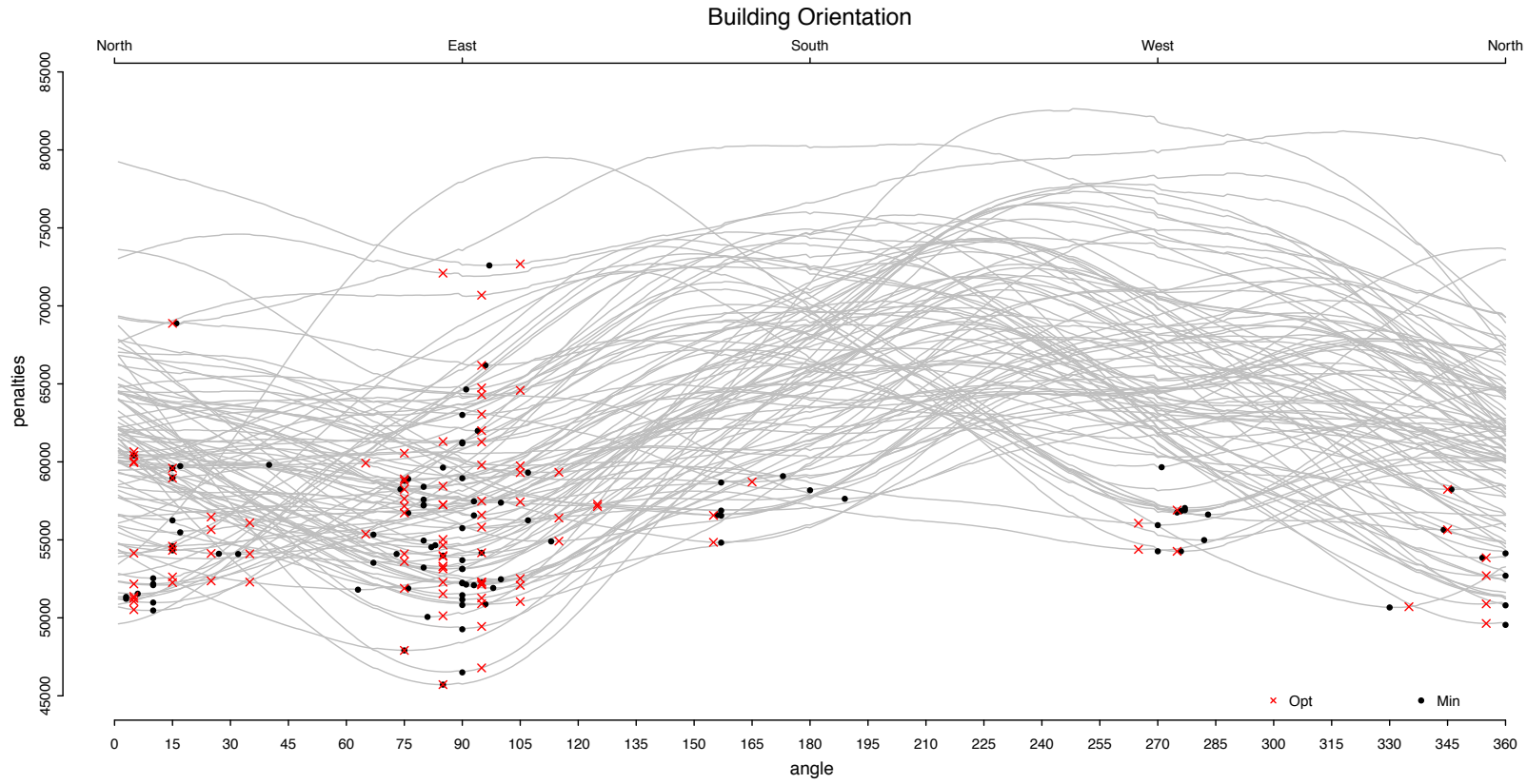

Figure 4: Building orientation curves with corresponding minimum value and optimization determined value.

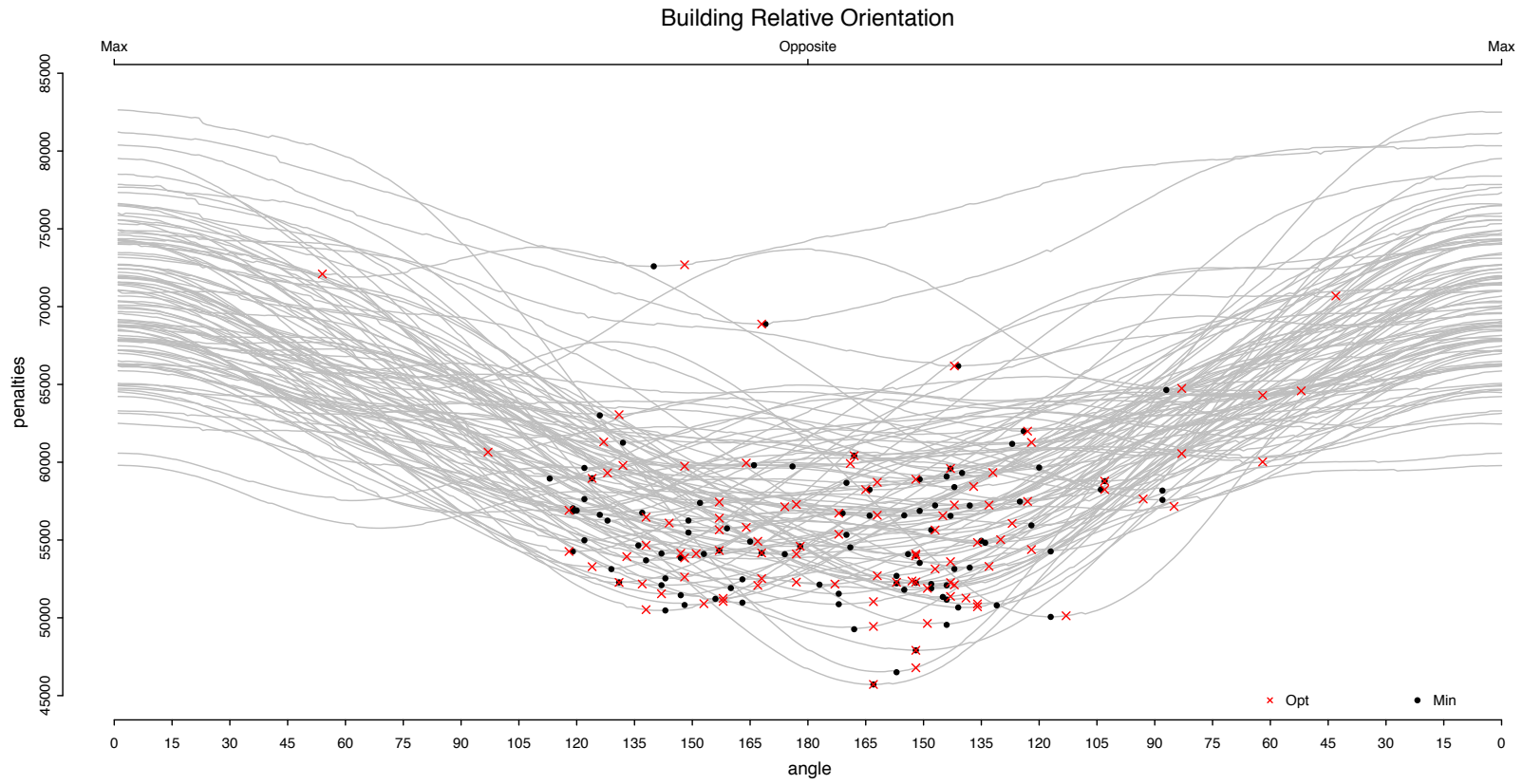

Figure 5: Building orientation curves adjusted according to the maximum value.

Three indicators were used to analyze the effectiveness of the building orientation operator. They are the difference in penalties in relation to the global optimum (1), the angle of deviation in relation to the global optimum (2), and finally, the percentage of error relatively to the curve amplitude (3).

The results for the indicators may be seen in Figure 6. The average difference in penalties in relation to the global optimum (1) was approximately 50, meaning that 50 degree-hours of air temperature divided by all the spaces of the floor plan, and by the 8760 hours of the year, the average temperature difference would be 0.0006ºC of improvement (top box plot).

The second indicator (2) determines the average angle of deviation in relation to the optimum angle (middle box plot of the Figure 6). The average deviation in this case was 5 degrees and ranged from 0 to 15 degrees.

Some designs have a flat building orientation curve, meaning that the deviation angle is not the best indicator of the algorithm success. For this reason, the last indicator (3) assesses the percentage of error in relation to the amplitude of the building orientation curve. In previous tests, the average percentage of error was approximately 0.4%. This is less than half a percent and it is below the previously stated threshold of admissible errors of 2%. The range of this indicator varied from 0% up to 2.4%.

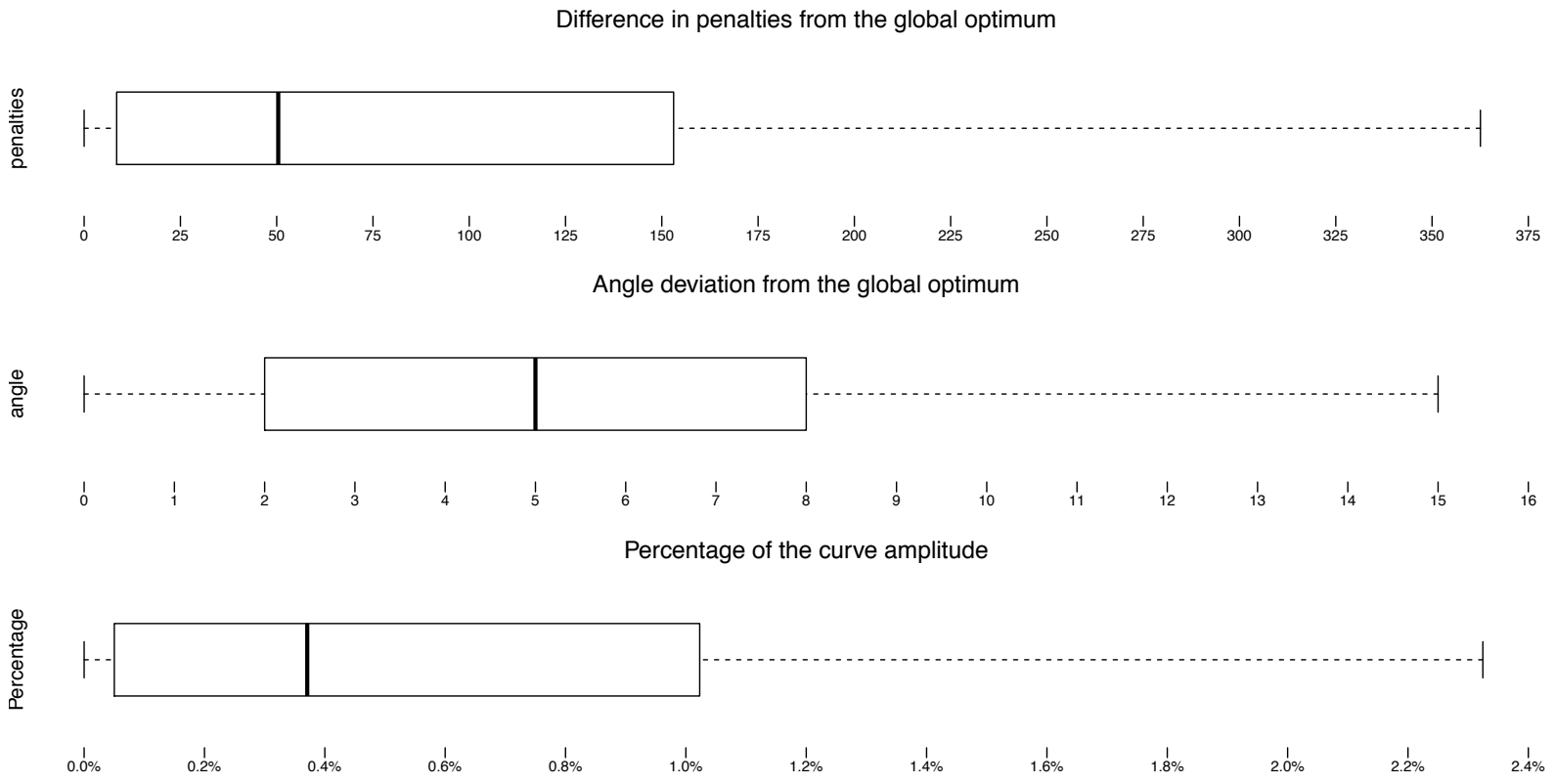

Figure 6: Building orientation operator success indicators.

## 4. Conclusion

In the approach presented in this paper, dynamic simulation of buildings was effectively used together with an optimization procedure, allowing the designs to be significantly improved. The use of an orientation variable was also included in a building energy optimization procedure.

The capability of the FPOP module to improve a previously selected set of floor plans according to thermal performance criteria equips the practitioner with thermal information on their preferences and constraints and the potential of each solution.

The building orientation is one of the design variables used in the optimization of the FPOP module. A building orientation operator must take two objectives into account, the correct solar orientation that minimizes thermal discomfort penalties and it must allow other design variable operators to fully reach their purpose. Each floor plan design has its unique building orientation performance curve, meaning that any change in the architectural elements such as walls, openings, spaces dimensions, and materials alter the profile of the curve.

The proposed technique that was employed such as the building orientation operator was tested for its effectiveness and efficiency using 100 floor plan designs with the same design program for the same location. The test has shown that the deviation from the global optimum orientation falls within the admissible error margin. Only 2.7% of all possible runs of the dynamic simulation were necessary to find a near optimum building orientation.